\documentclass[twocolumn, amssymb, aps, preprintnumbers,
amsmath,floatfix]{revtex4}

\usepackage{epstopdf}
\usepackage{graphics}
\usepackage{graphicx}
\usepackage{dcolumn}
\usepackage{bm}
\usepackage{longtable}
\usepackage{epsfig}
\usepackage{times}
\usepackage{url}
\usepackage{color}

\begin{document}
\title{Gravitational and Relativistic Deflection of X-Ray Superradiance}

\author{Wen-Te \surname{Liao}}
\email{wen-te.liao@mpi-hd.mpg.de}

\author{Sven \surname{Ahrens}}

\affiliation{Max-Planck-Institut f\"ur Kernphysik, Saupfercheckweg 1, D-69117 Heidelberg, Germany}
\date{\today}
\begin{abstract}
Exploring Einstein's theories of relativity in quantum systems, for example by using atomic clocks at high speeds can deepen our knowledge in physics. However, many challenges still remain on finding novel methods for detecting  effects of gravity and of special relativity and their roles in light-matter interaction. Here we introduce a scheme of x-ray quantum optics that allows for a millimeter scale investigation of the relativistic redshift by directly probing a fixed nuclear crystal in Earth's gravitational field with x-rays. 
Alternatively, a compact rotating crystal can be used to force interacting x-rays to experience inhomogeneous clock tick rates in a crystal. We find that an association of gravitational or special-relativistic time dilation with quantum interference will be manifested by  deflections of x-ray photons. Our protocol suggests a new and feasible tabletop solution for probing effects of gravity and special relativity in the quantum world.
\end{abstract}
\maketitle
Modern x-ray science \cite{xfel,Roehlsberger2004,Shvydko2004,Shvyd2011} opens an entirely new era of quantum optics \cite{Adams2013}. 
The vast and unexplored land of x-ray quantum optics \cite{Glover2010,Adams2013,Rohringer2012,Vagizov2014} provides many possibilities for both fundamental research \cite{Rohlsberger2010,Rohlsberger2012,Heeg2013} and applications \cite{Liao2012a,Vagizov2014,Cavaletto2014}. In particular, studies of the interaction between x-rays and nuclear condensed matter systems lead to novel control of x-ray photons \cite{Shvydko1996,Rohlsberger2012,Liao2012a,Vagizov2014}.
Furthermore, the well-known Pound-Rebka experiment \cite{Pound1960} shows that the M\"ossbauer effect is remarkably useful for the testing gravitational redshift \cite{Einstein1905,Einstein1915} by detecting the influence of Earth's gravity on propagating x-rays. Another result was obtained with a rotating nuclear crystal \cite{Hay1960}, demonstrating that the interaction between x-rays and nuclei is also useful for testing special relativity.
Therefore, x-ray quantum optics with nuclei may offer a new access to the exploration of gravity and special relativity in quantum systems. 
The gravitational redshift indicates that clocks at high altitudes run faster than those at sea level under the influence of Earth's gravity \cite{Chou2010}. 
Also, according to special relativity, a moving clock ticks at a slower rate than a fixed one in an observer's rest frame \cite{Ives1938,Reinhardt2007}. 
Thus, within a crystal, identical particles at different lattice sites form an array of clocks which tick at inhomogeneous rates, when they are influenced by gravity or by time dilation of an inhomogeneous motion. This opens the question of how this compact configuration affects the quantum mechanical coherence of such a crystal.

\begin{figure}[b]
\vspace{-0.4cm}
  \includegraphics[width=0.48\textwidth]{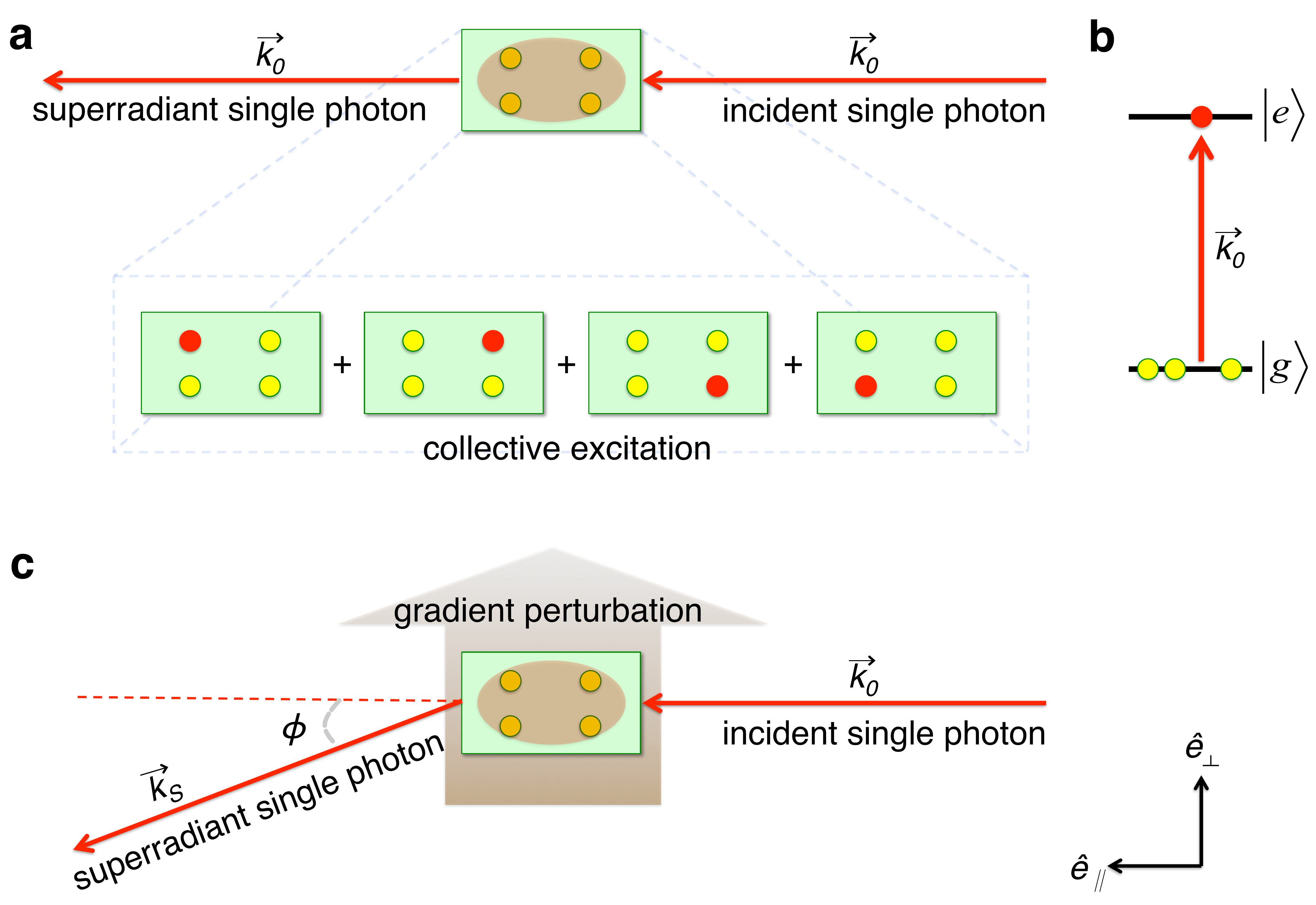}
  \caption{\label{fig1}
\textbf{Superradiant single photon.}
\textbf{a,}  
An incident single photon with wave vector $\vec{k}_{0}$ is absorbed and shared by an ensemble of two-level quantum particles. Without knowing which one is excited, a collective excitation, which is a superposition of all possibilities, will form. The decay of such a delocalized collective excitation will be followed by a directional emission of a single photon along the incident $\vec{k}_{0}$. This coherently reemitted photon is called superradiant single photon. Yellow (red) dots illustrate particles in the ground (excited) state, and red arrows depict both the incident and the reemitted single photon. 
\textbf{b,}
A collective excitation is created in a collection of two-level particles while a particle is excited from its ground level $\vert g \rangle$ to a higher energy level $\vert e \rangle$ by absorbing a single photon with wave vector $\vec{k}_{0}$.
\textbf{c,}
A superradiant single photon is deflected by a gradient perturbation depicted by the brown upward arrow. $\phi$ is the deflection angle between incident wave vector $\vec{k}_{0}$ and deflected wave vector $\vec{k}_{S}$.
  }
\end{figure}

Here, we investigate a scenario that manifests distinct deflections of a single photon under the action of gravity or special relativity and quantum interference.
Considering long-lived nuclear states,
one could shine an x-ray on a thin nuclear crystal to create a collective excitation that is simultaneously perturbed by relativistic time dilatation. 
Since the excitation is delocalized over the whole ensemble of nuclei \cite{Roehlsberger2004, Scully2006, Rohlsberger2010}, it evolves with an inhomogeneous rate caused by Earth's gravity.
We find that the inhomogeneous evolution of a delocalized excitation causes a deflection of the reemitted single photon. 
This time-dependent deflection suggests that the photon trajectory can be influenced by Earth's gravity even though it was stored as a stationary quantum excitation in a crystal.
An analogue with slow light propagation of about 100\,m/s in media proposes a light deflection of around $10^{-9}$ degrees \cite{Dressel2009}, being challenging for detections in the optical domain.

A quantum collective excitation occurs when a single photon is absorbed by a collection of $N$ particles. This single photon is then shared by $N$ particles, that leads to a delocalized collective excitation state as depicted in Fig.~\ref{fig1}~a.
The collective excitation state can be written as \cite{Roehlsberger2004, Scully2006, Rohlsberger2010}
\begin{equation}
\vert E \rangle=\frac{1}{\sqrt{N}}\sum_{\ell}^{N}e^{i \vec{k}_{0} \cdot\vec{r}_{\ell}}e^{i (\nu_{\vec{k}_{0}}-\omega_{\ell})t}\vert g\rangle\vert e_{\ell}\rangle.
\label{eq1}
\end{equation}
Here $\vert g\rangle\vert e_{\ell}\rangle$ denotes that particle $\ell$ at position $\vec{r}_{\ell}$ is in its excited state $\vert e\rangle$, while the other $N-1$ particles remain in the ground state $\vert g\rangle$ as illustrated in Fig.~\ref{fig1}~b. The excitation energy of particle $\ell$ is $\hbar\omega_{\ell}$, where $\hbar$ is the reduced Planck constant.
Furthermore, $\nu_{\vec{k}_{0}}$ and $\vec{k}_{0}$ are the angular frequency and wave vector of the incident single photon, respectively. As a result of quantum interference between the emission from each crystal site, a directional reemission of a single photon, namely, superradiance, follows the decay of state $\vert E \rangle$ along the direction $\vec{k}_{0}$ of an incident photon \cite{Rohlsberger2010}. The directional superadiance is routinely observed in nuclear forward scattering of x-rays \cite{Rohlsberger2010} using a nuclear solid-state crystal that is typically few microns thick and few millimeters in diameter \cite{Shvydko1996,Roehlsberger2004}. These tiny dimensions therefore give the upper bound of the spatial scale where relativistic effects could be probed by our scheme.

A gradient perturbation, as depicted in Fig.~\ref{fig1}~c, can be utilized to control the direction of superadiance.
As some external gradient perturbation is applied to the whole ensemble \cite{Karpa2006, Adams2013}, the originally constant $\omega_{\ell}$ becomes inhomogeneous $\omega_{\ell}(\vec{R}+\vec{r}_{\ell})$, where $\vec{R}$ is the position of the ensemble relative to the origin of the perturbation and $\vec{r}_{\ell}$ is the particle position in the ensemble coordinate.
Assuming that the ensemble size is much smaller than $|\vec{R}|$, one can use the Taylor expansion 
$\omega_{\ell}(\vec{R}+\vec{r}_{\ell})\simeq \omega_{\ell}(\vec{R})+\nabla\omega_{\ell}(\vec{R})\cdot\vec{r}_{\ell}$.
By substituting the expanded $\omega_{\ell}(\vec{R}+\vec{r}_{\ell})$ into equation~\eqref{eq1}, the collective excitation state becomes (see Methods and Supplementary Information)
\begin{equation}
\vert E \rangle=\frac{1}{\sqrt{N}}\sum_{\ell}^{N}e^{i \vec{k}_{S}(t) \cdot\vec{r}_{\ell}}e^{i [\nu_{\vec{k}_{0}}-\omega_{\ell}(\vec{R})]t}\vert g\rangle\vert e_{\ell}\rangle,
\label{eq2}
\end{equation}
where the new time-dependent wave vector $\vec{k}_{S}(t)=\vec{k}_{0}-t\nabla\omega_{\ell}(\vec{R})$ indicates the  deflection of superadiance. If $\vec{k}_0$ and $\nabla\omega_{\ell}(\vec{R})$ are perpendicular to each other, the deflection angle $\phi(t)$ between $\vec{k}_0$ and $\vec{k}_{S}(t)$ can be written as
\begin{equation}
\phi(t) = \tan^{-1}\left[\frac{|\nabla\omega_{\ell}(\vec{R})|t}{|\vec{k}_{0}|}\right]\,.
\label{eq3}
\end{equation}
In what follows, we show that the above superradiant deflection described by equations~(\ref{eq2}-\ref{eq3}) may be introduced by Earth's gravity and by the time dilation of special relativity in a rotating system.

\begin{figure*}
\begin{center}
\vspace{-0.4cm}
  \includegraphics[width=0.7\textwidth]{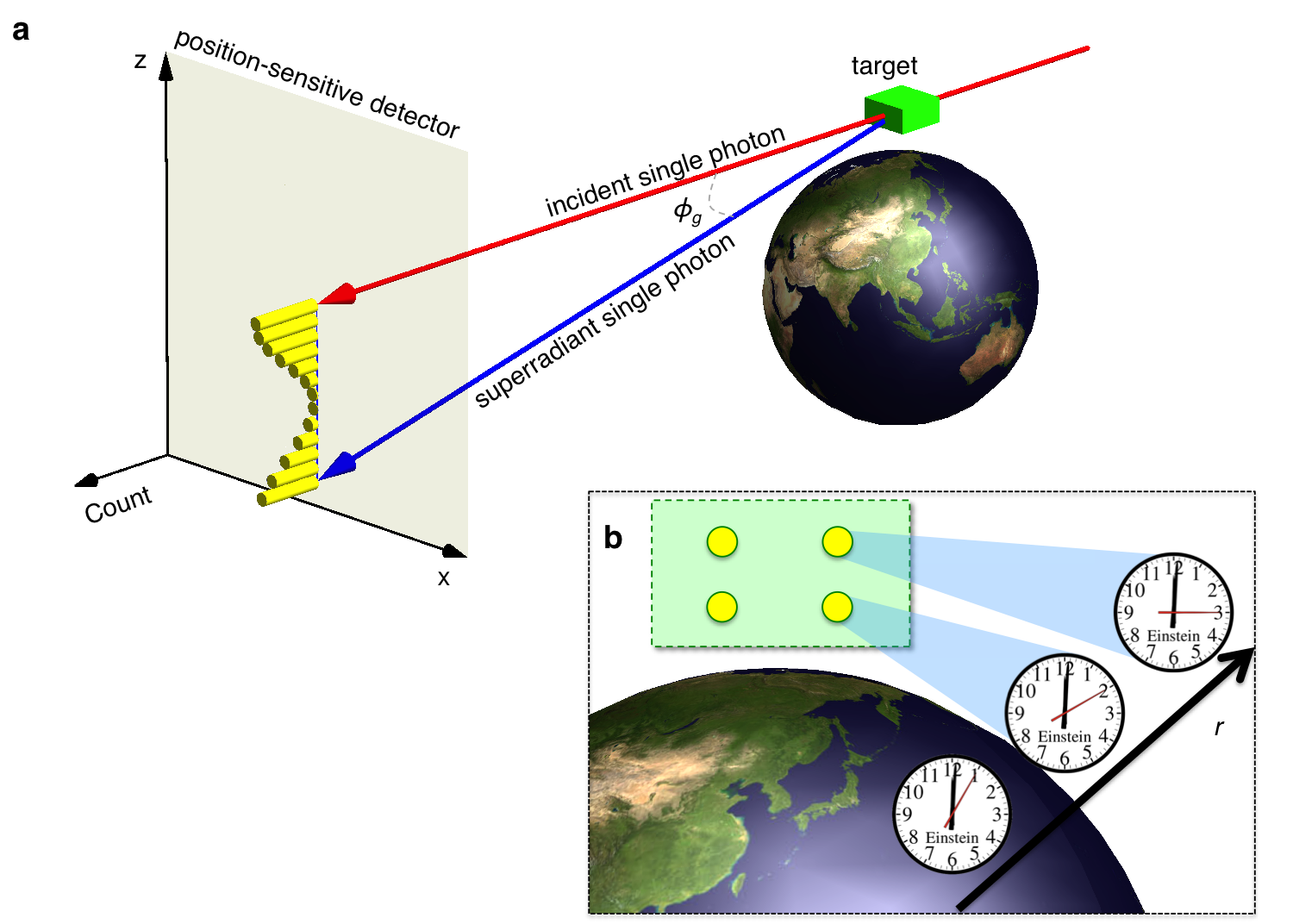}
  \caption{\label{fig2}
\textbf{Gravitational deflection of x-ray superradiance.} 
\textbf{a,} 
Earth's gravity deflects the superradiant single x-ray photon from its incident direction with a time-dependent angle $\phi_{g}$. 
The green cuboid depicts a fixed nuclear crystal in a gravitational field of the Earth, the red (blue) arrow represents the incident (reemitted) single photon and yellow horizontal bars illustrate the count numbers at different positions of a position-sensitive detector.
\textbf{b,}
Influence of gravity on proper time: Clocks closer to the earth run slower. The same happens to fixed nuclei (yellow dots) in a crystal (depicted by the green rectangle). Earth, crystal size and clock ticks are not on scale.
  } 
\vspace{0.1cm}
  \includegraphics[width=0.7\textwidth]{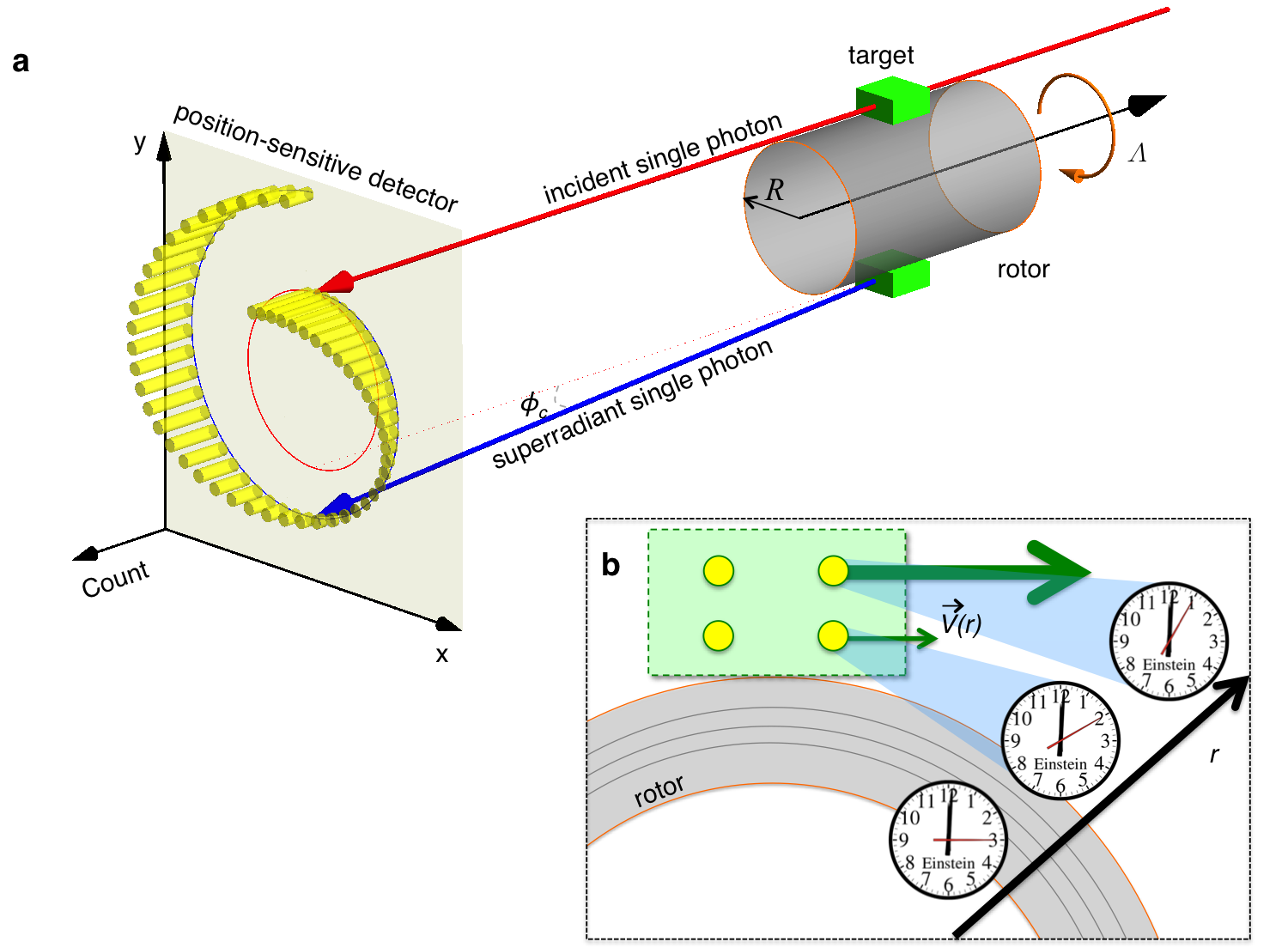}
  \caption{\label{fig3}
\textbf{X-ray superradiance under the influence of special relativity.}
\textbf{a,}  
X-ray photon deflection by time dilation of special relativity, where the crystal (green cuboid) is attached to a rotor with radius $R$ and is subject to a rotation with angular frequency $\varLambda$. 
Under the influence of the inhomogeneous relativistic time dilation,
the incident single x-ray photon (red arrow) is bent along the blue arrow with a time-dependent angle $\phi_{c}$ and time-dependent reemission probability (yellow bars on the position-sensitive detector). The red circle on the detector illustrates the trajectory of the registered signals without considering the inhomogeneous relativistic time dilation.
\textbf{b,}
According to the time dilation effect, the tick rate of the proper time of each nucleus at different radii $r$ depends on the local velocity $\vert\vec{V}(r)\vert=\varLambda r$ within a crystal.
Also, green arrows illustrate local velocities of nuclei (yellow dots) at different sites of a rotating crystal (green rectangle).
  }    
\end{center}
\end{figure*}

\begin{table}[b]
\vspace{-0.4cm}
\caption{\label{table1}
$\phi_{g}(\tau_{coh})$ and $\phi_{c}(\tau_{coh})$ are the maximum deflection angles of superradiant photons induced by Earth's gravity and the inhomogeneous speed of a rotor, respectively. 
Here $\tau_{coh}$ is the coherence time of the corresponding nuclear transition. The parameters of the rotor are $R$ = 5 mm and $\varLambda$ = $2\pi\times$ 70 kHz. Also, $E_{e}$ is the nuclear excited state energy, which also corresponds to the used photon energy \cite{NSDD}.
}
\center{
\begin{tabular}{cccccc}
\hline
crystal                     & $E_{e}$                & coherence time      & $\phi_{g}(\tau_{coh})$   & $\phi_{c}(\tau_{coh})$      \\
                           & (keV)                  &  $\tau_{coh}$       & (degrees)                 & (degrees)                    \\ 
\hline 
$^{45}$Sc                  & 12.4                   &459      ms          &  $8.6\times 10^{-7}$     & $90$                        \\ 
$^{57}$Fe                  & 14.41                  &141      ns          &  $2.6\times 10^{-13}$    & $2.6\times 10^{-5}$         \\ 
$^{67}$Zn                  & 93.31                  &13.09    $\mu$s      &  $2.5\times 10^{-11}$    & $2.4\times 10^{-3}$         \\ 
$^{73}$Ge                  & 13.28                  &4.21     $\mu$s      &  $7.9\times 10^{-12}$    & $7.8\times 10^{-4}$         \\ 
$^{109}$Ag                 & 88.03                  &57.13         s      &  $1.1\times 10^{-4}$     & $90$                        \\  
$^{181}$Ta                 & 6.24                   &8.73     $\mu$s      &  $1.6\times 10^{-11}$    & $1.6\times 10^{-3}$         \\ 
$^{182}$Ta                 & 16.27                  &408      ms          &  $7.7\times 10^{-7}$     & $90$                        \\ 
$^{229}$Th:CaF$_{2}$       & 0.0078                 &1        ms          &  $1.9\cdot 10^{-9}$      & $0.18$                      \\
\hline
\end{tabular}
}
\end{table}
The considered system on the surface of the Earth is depicted in Fig. \ref{fig2}. Trains of monochromatic x-ray synchrotron radiation pulses or single x-ray photons resonant to a nuclear transition impinge on the crystal along the red arrow \cite{Roehlsberger2004}. The x-ray pulses are spaced at intervals of the coherence time $\tau_{coh}$ defined by, e.g., the lifetime of a nuclear excited state.  Each weak synchrotron radiation pulse or single x-ray photon mostly excites a single nucleus in a crystal and so creates the above collective excitation state $\vert E \rangle$ defined by Eq.~(\ref{eq2}) \cite{Shvydko1996,Roehlsberger2004,Rohlsberger2010,Liao2012a}.
Because of gravity, particles at various crystal sites evolve differently in time, depending on their position. 
The gravitational redshift deduced from the Schwarzschild metric gives $|\nabla\omega_{\ell}(\vec{R})|=\omega \times 1.09\times 10^{-16} / \textrm{m}$ on Earth's surface, which is well consistent with the experimental redshift found by Pound and Rebka \cite{Pound1960}. 
Because of this inhomogeneous quantum phase evolution, the superradiant photons are reemitted along the blue arrow in Fig.~\ref{fig2}~a and deflected by a time-dependent angle (see Supplementary Information):
\begin{equation}
\phi_{g}(t)\approx\tan^{-1}\left[ \frac{G M_E t}{c r_E^2 \sqrt{1 - \frac{2 G M_E}{c^2 r_E}}} \right]\,, \nonumber
\label{eq4}
\end{equation}
where $G$ is the gravitational constant, $M_E$ is the mass of the Earth and $r_E$ is the Earth's radius. 
Since gravity is a weak force, the resulting angular deflection velocity is $\partial_{t}\phi_{g} \approx 1.9\times 10^{-6}$ degree/s, 
such that one needs $\tau_{coh}>0.5$ s to observe a resolvable deflection angle of $10^{-6}$ degrees  with modern x-ray optics.
As listed in Table \ref{table1}, some candidate transitions of nuclei, e.g., $^{45}$Sc, $^{109}$Ag and $^{182}$Ta, already give large enough $\phi_{g}$ to be observed.
This gravitational deflection depends on the strength of gravity and can be observed more significantly in the vicinity of astronomical objects like neutron stars or black holes. 
For instance, if the Earth was compressed to a compact size of 1.3 km in diameter, $\partial_{t}\phi_{g}$ would speed up to about 180 degree/s.
In what follows, we show that such strong deflection can be achieved by an inhomogeneous time dilatation of special relativity. 

We turn to the realization of a larger deflection of superradiance by a fast rotating crystal as demonstrated in Fig.~\ref{fig3}. An inomogeneous clock tick rate in a crystal caused by time dilation of special relativity mimics the time gradient of a gravitational field  \cite{Hay1960}.
A nuclear crystal is fixed on a rotor with a rotating angular frequency $\varLambda$ and a radius $R$. 
As the whole system rotates, a particle at site $r_{\ell}$ of the crystal moves at velocity $\vert\vec{V}(r_\ell)\vert=\varLambda (R+r_\ell)$, as illustrated in Fig.~\ref{fig3}~b.
According to special relativity \cite{Hay1960}, the transition angular frequency of a particle at site $r_{\ell}$ of a crystal becomes 
$\omega_\ell
=
\omega_{0}\sqrt{1-\left[ \varLambda\left( R+r_{\ell}\right)/c\right]^{2}}$,
and therefore a spatial gradient of $\omega_\ell$ points in the radial direction of the rotor in this configuration. Here, $\omega_{0}$ is the transition angular frequency of a particle in its rest frame.
For creating a collective excitation in a rotating crystal as depicted in Fig.~\ref{fig3}, the crystal is also illuminated by a train of x-rays spaced by $\tau_{coh}$  along the red arrow. Subsequently, single x-ray photons scattered off the crystal are registered by a position-sensitive detector along the blue arrow in Fig.~\ref{fig3}~a with a time-dependent deflection angle (see Supplementary Information):
\begin{equation}
\phi_{c}(t)\approx\tan^{-1}\left[ \frac{R \varLambda^2 t}{\sqrt{c^{2}- \varLambda^2 R^2}}\right]. \nonumber
\label{eq5}
\end{equation}
A rotor with a radius of 5 mm and rotation frequency of 70 kHz, previously used in the so-called nuclear lighthouse effect \cite{Roehlsberger2004}, allows for an angular velocity $\partial_{t}\phi_c \approx 185$ degree/s, yielding distinct deflection angles with the need of only $\tau_{coh}>10$ ns. 
As demonstrated in Table~\ref{table1}, many nuclear transitions, e.g., $^{57}$Fe and $^{67}$Zn, are suitable for that purpose. Remarkably, some transitions with $\tau_{coh}>0.31$ s may even allow $\phi_c=90$ degrees and deliver an unmistakable signature of relativistic deflections.
We emphasize that the key parameter for the present effect is the coherence time of an interacting transition  with photons. 
Coherence times of few milliseconds, 1 s and 40 s have been proposed or experimentally achieved in systems of $^{229}$Th:CaF$_{2}$ \cite{Kazakov2012}, diamond nitrogen-vacancy center \cite{Maurer2012} and  Pr$^{3+}$:Y$_{2}$SiO$_{5}$ \cite{Heinze2013}, respectively. These quantum memories together with recent diamond silicon-vacancy center \cite{Neu2011} deserve attention and may be suitable for our scheme at different energy scales.
Moreover, a semiclassical analysis shows that the deflection angle can be enhanced by a factor of optical depth (see Supplementary Information), and so using an optically dense crystal may result in larger deflections.

In conclusion, we have put forward a scheme to probe effects of gravity and special relativity on a compact millimeter scale via x-ray quantum optics. The underlying effect
presented here may also generically occur on different space and time scales.
For example, bending superadiance may also happen on an astronomical scale in a gravitational field superimposed by interstellar dust and gas clouds. 
In contrast to gravitational lensing \cite{Schneider2006}, deflected superradiance would cause a time-dependent distortion of images of astronomical objects. This may provide a novel way for exploring light-matter interactions with the theory of relativity on a large scale.
We thank C. H. Keitel, A. P\'allfy, Z. Harman, E. Yakaboylu, S. M. Cavaletto, C. O'Brien, K. Heeg, M. G\"arttner, T. Babinec, P. Hemmer and A. Svidzinsky for valuable discussions.


\pagebreak
\widetext
\begin{center}
\textbf{\large Supplemental Materials: Gravitational and Relativistic Deflection of X-Ray Superradiance}

{\normalsize Wen-Te Liao and Sven Ahrens}

Max-Planck-Institut f\"ur Kernphysik, Saupfercheckweg 1, 69117 Heidelberg, Germany

(Dated: \today)
\end{center}
\setcounter{equation}{0}
\setcounter{figure}{0}
\setcounter{table}{0}
\setcounter{page}{1}
\makeatletter
\renewcommand{\theequation}{S\arabic{equation}}
\renewcommand{\thefigure}{S\arabic{figure}}
\renewcommand{\bibnumfmt}[1]{[S#1]}
\renewcommand{\citenumfont}[1]{S#1}

\begin{longtable}{cl}
\caption{\label{parametertable}
Symbol table. 
}\\
\hline
Symbol                     & Explanation                                                                          \\
\hline                                                                                                            \\
$\vert g \rangle$          & ground state of a particle                                                           \\
$\vert e_{\ell}\rangle$    & the $\ell$th particle is in its excited state                                        \\
$\vert 0 \rangle$          & vacuum state                                                                         \\
$\vert 1\rangle_{\vec{u}}$ & one photon Fock state of mode $\vec{u}$                                              \\
$\vert E \rangle$          & collective excitation state                                                          \\
$\vert \psi \rangle$       & collective polariton state                                                           \\
$t$                        & time in the lab frame                                                                \\
$\omega$                   & transition angular frequency in the rest frame of a particle                         \\
$\omega'$                  & transition angular frequency in the lab frame                                        \\
$\nu_{\vec{k}}$            & photon angular frequency $c|\vec{k}|$                                                \\
$\vec{r}_{\ell}$           & position vector of the $\ell$th particle                                             \\
$\vec{k}_{0}$              & wave vector of an incident single photon                                             \\
$\Theta$                   & azimuth angle of the cylindrical coordinates                                         \\
$k_{0}$                    & $|\vec{k}_{0}|$                                                                      \\
$\vec{k}_{S}$              & wave vector of the remitted superradiant single photon                               \\
$\hat{e}_{\parallel}$      & unit vector that is parallel to $\vec{k}_{0}$                                        \\
$\hat{e}_{\perp}$          & unit vector that is perpendicular to $\vec{k}_{0}$                                   \\
$\widehat{a}^{\dagger}_{\vec{k}}$& photon creation operator for some specific $\vec{k}$                           \\
$\widehat{a}_{\vec{k}}$    & photon annihilation operator for some specific $\vec{k}$                             \\
$g_{\vec{k}}$              & particle-field coupling rate for some $\vec{k}$ mode photon                          \\
$c$                        & speed of light                                                                       \\
$\widehat{\sigma}_{+}^{j}$ & raising operator for particle at position $\vec{r}_{j}$                              \\
$\widehat{\sigma}_{-}^{j}$ & lowering operator for particle at position $\vec{r}_{j}$                             \\
$\hbar$                    & reduced Planck constant                                                              \\
$R$                        & radius of a rotor                                                                    \\
$\varLambda$               & rotating angular frequency of a rotor                                                \\
$\theta$                   & polar angle of the earth spherical coordinates                                       \\
$\varphi$                  & azimuth angle of the earth spherical coordinates                                     \\
$r_E$                      & Earth radius, 6371 $km$                                                              \\
$r_s$                      & Schwarzschild radius                                                                 \\
$F$                        & Schwarzschild function                                                               \\
$G$                        & gravitational constant, 6.67384$\times 10^{-11} m^3 kg^{-1} s^{-2}$                  \\
$M$                        & mass of a gravitational source, for the Earth $M_{\bigoplus}=5.97219\times 10^{24} kg$ \\ 
$Q$                        & $\nabla\omega'\cdot\hat{e}_{\perp}$                                                  \\ 
$\beta(t)$                 & $\nabla\omega'\cdot\hat{e}_{\parallel}$                                              \\ 
$P(t)$                     & $\int_{0}^{t}\beta(t)dt$                                                             \\
$A(t)$                     & time derivative of Zeeman field gradient                                             \\
$\phi(t)$                  & deflection angle of the remitted superradiant single photon                          \\
$\phi_{g}(t)$              & deflection angle of the remitted superradiant single photon in gravity               \\
$\phi_{c}(t)$              & deflection angle of the remitted superradiant single photon  in a rotating crystal
\\
$\mu_{n}$                  & particle's magnetic dipole moment                                                    \\
$\Omega_{p}$               & Rabi-frequency of probe field                                                        \\
$\wp$                      & particle number density in a crystal                                                 \\
$L$                        & crystal length                                                                       \\
$\sigma$                   & on-resonance cross section of the considered transition                              \\
$\vec{d}$                  & transition dipole moment                                                             \\
\hline
\end{longtable}

The central purpose of this supplemental material is to show the deflection of scattered photons off an ensemble of particles caused by the Earth's gravity or by an inhomogeneous time dilation of special relativity. As illustrated in Fig.~1 and Fig.~2, 
a train of single photons impinges on the target to initially create the collective excitation state expressed by Eq.~(\ref{Fe_exciton2}). A series of scattered single photons will then follow the coherent decay of the exciton and be registered by a position-sensitive detector. Let's begin with the description of our system.
The collective excitation state of $N$ particles in the cylindrical coordinate can be written as \cite{Duan2001, roehlsberger2004, scully2006, Riedmatten2008, svidzinsky2008}
\begin{equation}
\vert E \rangle=\frac{1}{\sqrt{N}}\sum_{\ell}D_{\ell}(t,\vec{r}_{\ell})e^{i \vec{k}_{0} \cdot\vec{r}_{\ell}}\vert g\rangle\vert e_{\ell}\rangle,
\label{Fe_exciton2}
\end{equation}
where $\vert g\rangle\vert e_{\ell}\rangle$ denotes that only the $\ell$th particle at position $\vec{r}_{\ell}=(r_{\ell}, z_{\ell})$ is excited  and all remaining $N-1$ particles stay in the ground state. Furthermore, $\vec{k}_{0}$ is the wave vector of the incident resonant photon and coefficient $D_{\ell}(t,\vec{r}_{\ell})$ is caused by the $\vec{r}_{\ell}$-dependent and nonuniform environment. 
The complete collective polariton state is
\begin{equation}\label{Fe_totalstate2}
\vert\psi\rangle=\vert G\rangle\otimes\sum_{\vec{u}}B_{\vec{u}}(t)\vert 1\rangle_{\vec{u}}+\frac{1}{\sqrt{N}}\sum_{\ell}D_{\ell}(t,\vec{r}_{\ell})e^{i \vec{k}_{0} \cdot\vec{r}_{\ell}}\vert g\rangle\vert e_{\ell}\rangle\otimes\vert 0\rangle,
\end{equation}
with the initial condition
\begin{equation}\label{ic}
B_{\vec{u}}(0)=0, \; D_{\ell}(0)=1, \; \vec{k}_{0}=k_{0} \widehat{z},
\end{equation}
where $\vert G\rangle$ denotes all particles are in the ground state. Moreover, $\vert 1\rangle_{\vec{u}}$ and $\vert 0\rangle$ denote the one photon Fock state of wave vector $\vec{u}$ and the vacuum state, respectively. The combined photon-particle system can be described by the Schr\"odinger equation
\begin{equation}\label{Fe_SE2_rest}
\partial_{t}\vert\psi\rangle=-\frac{i}{\hbar}\widehat{H}\vert\psi\rangle\,,
\end{equation}
where the Hamiltonian is
\begin{align}\label{Fe_JCmodel2}
\widehat{H}=
\hbar\sum_{\vec{k}}\nu_{\vec{k}}\widehat{a}^{\dagger}_{\vec{k}}\widehat{a}_{\vec{k}}
+\hbar\sum_{j}\left[ \omega + \Delta(r_{j},z_{j},t) \right] \vert e_{j}\rangle\langle e_{j}\vert
+\hbar \sum_{j,\vec{k}}g_{\vec{k}}\left[ e^{i\vec{k}\cdot \vec{r}_{j}}\widehat{\sigma}_{+}^{j}\widehat{a}_{\vec{k}}+
e^{-i\vec{k}\cdot \vec{r}_{j}}\widehat{\sigma}_{-}^{j}\widehat{a}^{\dagger}_{\vec{k}}\right]\,.
\end{align}
Here, base on the experimental result given by R. V. Pound and G. A. Rebka \cite{Pound1960}, and also another one given by H. J. Hay et al. \cite{hay1960}, the effect of relativistic time dilation can be treated as an additional energy shift
\begin{equation}
\label{rotatek}
\Delta(\vec{r}_{j},t) = \alpha+\left[Q \widehat{e}_{\perp}+\beta(t)\widehat{e}_{\parallel}\right]\cdot\vec{r}_{j},
\end{equation}
where the parameters $\alpha$ and $Q$ are determined by gravity and the inhomogeneous speed assumed to be perpendicular to $\vec{k}_{0}$ (see sections \ref{sec:general_relativity} and \ref{sec:rotating_sample}, respectively).
Additionally, a dynamical Stark or Zeeman shift term $\beta(t)\widehat{e}_{\parallel}\cdot\vec{r}_{j}$ is applied \cite{adams2013} along the direction of $\vec{k}_{0}$ to fulfill the energy-momentum conservation (see section \ref{semc}).
Moreover, ($\widehat{e}_{\perp}$, $\widehat{e}_{\parallel}$) is the unit vector (perpendicular, parallel) to $\vec{k}_{0}$, and $\widehat{e}_{\perp}$ is parallel to the direction of gravity or that of rotor's radius in what follows.
Also, $\vec{k}$ and $\nu_{\vec{k}}$ are the wave vector and the angular frequency, respectively, of the interacting photon. $(\widehat{\sigma}_{+}^{j},\widehat{\sigma}_{-}^{j})$ are the particle (raising, lowering) operators at position $\vec{r}_{\ell}$, and $(\widehat{a}^{\dagger}_{\vec{k}},\widehat{a}_{\vec{k}})$ are the photon (creation, annihilation) operators for a specific $\vec{k}$. $\vert 1\rangle_{\vec{k}}$ and $\vert 0\rangle$ denote the one photon Fock state of wave vector $\vec{k}$ and the vacuum state, respectively, and $g_{\vec{k}}$ is the particle-field coupling rate. Furthermore, the energy of each particle's ground state and that of excited state are assumed to be zero and $\hbar \omega$, respectively.
In what follows, we derive the form of $\alpha$, $Q$ and the required $\beta(t)$.

\section{A fixed sample in the Earth's gravity}
\label{sec:general_relativity}
In general relativity, the time at infinity is related to the proper time by the line element of the Schwarzschild metric \cite{schwarzschild,gravitationbook}. We use the Schwarzschild metric  of a black hole here, because it is identical with the metric of the earth outside of the earth and has a simple analytic form. The line element of the Schwarzschild metric is
\begin{equation}
 c^2 d t'^2 = c^2 F(r) dt^2 - F(r)^{-1} dr^2 - r^2 d \theta^2 - r^2 \sin^2 \theta d \varphi^2 \,,\label{eq:lineelement}
\end{equation}
where $F(r)$ is the Schwarzschild function
\begin{equation}
 F(r) = 1 - \frac{r_s}{r}\,,
\end{equation}
with the Schwarzschild radius of the black hole event horizon
\begin{equation}
 r_s = \frac{2 G M}{c^2}
\end{equation}
and the gravitational constant $G$ and the mass $M$. $M$ can either be the mass of a black hole or, as in our case, the mass of the earth. At fixed radius $r$ the proper time $t'$ and the time $t$ at infinity are related by
\begin{equation}
 d t' = \sqrt{F(r)} dt \,.
\end{equation}
Since the transition frequencies are inversely proportional to the infinitesimal time element one gets
\begin{equation}
 \omega' = \omega \sqrt{F(r)} \simeq \omega \left[ \sqrt{F(r_E)} + \frac{r_s}{2 r_E \sqrt{F(r_E)}} \frac{r_\ell}{r_E} + \mathcal{O} \left(r_{\ell}^2\right) \right].
\end{equation}
One can then define two parameters
\begin{subequations}
\label{eq:alpha_and_Q_GR}
\begin{align}
\alpha &= \omega\left[ \sqrt{F(r_E)}-1\right]  \,, \\
Q &= \nabla \omega'\cdot\widehat{e}_{\perp}=\frac{\omega r_s}{2 r_E^2 \sqrt{F(r_E)}} \label{Qcondition_GR}\,.
\end{align}
\end{subequations}
The time, with respect to which the system is described within this ansatz is the time at the asymptotic flat space-time, infinitely far away from the earth at which $F(r)=1$. The parameter $\alpha$ accounts for the additional phase aquired by the superradiant crystal's finite radial position $r_E$ at earth surface, wereas the parameter $Q$ accounts for an additional height dependent phase change caused by gravity within the crystal.

\section{A rotating sample} 
\label{sec:rotating_sample}
A solid state sample is attached on a rotor with a rotating angular frequency $\varLambda$ and a radius $R$.
According to time dilation in special relativity:
\begin{eqnarray}\label{TD2}
\omega'
&=&
\omega\sqrt{1-\left[ \frac{\varLambda\left( R+r_{\ell}\right) }{c}\right]^{2}}
\nonumber \\
&\simeq &
\omega\left\lbrace 
\sqrt{1-\left( \frac{\varLambda R }{c}\right)^{2}}
-\frac{\frac{\varLambda^{2}R^{2}}{c^{2}}\left(\frac{r_{\ell}}{R}\right)}{\sqrt{1-\left( \frac{\varLambda R }{c}\right)^{2}}}-\mathcal{O}\left[\left(\frac{r_{\ell}}{R}\right)^2\right]
\right\rbrace, 
\end{eqnarray}
where $\varLambda$ is the angular frequency of the used rotor, $c$ is the speed of light, $R$ is the radius of the used rotor and $R+r_{\ell}$ is radial position of ${\ell}$th particle. $\omega$ and $\omega'$ are the particle's transition angular frequencies in the rest frame of each particle and the lab frame, respectively. One could identify\
\begin{subequations}
\label{eq:alpha_and_Q}
\begin{eqnarray}
\alpha &=& \omega\left[  \sqrt{1- \frac{\varLambda^2 R^2 }{c^2}}-1\right] \,, \\
Q &=&\nabla \omega'\cdot\widehat{e}_{\perp}=-\frac{1}{c^2}\frac{R \varLambda^2 \omega}{\sqrt{1- \frac{\varLambda^2 R^2}{c^2}}} \label{Qcondition}\,.
\end{eqnarray}
\end{subequations}
Within the ansatz of this rotating system, the reference time is that of the non-rotating lab. The parameter $\alpha$ introduces a time-dependent phase caused by special relativity of the rotating sample at radius $R$. The parameter $Q$ accounts for the additional time-dilation, caused by the different velocity which each nucleus with distance $R + r_l$ has.

\section{Deflection of interacting photons}
To investigate the dynamics of our system, the Weisskopf-Wigner theory \cite{scully1997} with $\partial_{t}\vert\psi\rangle=-\frac{i}{\hbar}\widehat{H}\vert\psi\rangle$ is used \cite{svidzinsky2008,scully1997}. We obtain the equation of motion for $B_{\vec{u}}$ and $D_{\ell}$:
\begin{equation}\label{Fe_Ck12}
\vert G\rangle\otimes\sum_{\vec{u}}\vert 1\rangle_{\vec{u}}\partial_{t}B_{\vec{u}}=
-i\left\lbrace 
\vert G\rangle\otimes\sum_{\vec{u}}\vert 1\rangle_{\vec{u}}\nu_{\vec{u}}B_{\vec{u}}
+\frac{1}{\sqrt{N}}\vert G\rangle\otimes\sum_{\ell,\vec{k}}
g_{\vec{k}}e^{i \left( \vec{k}_{0}-\vec{k}\right) \cdot \vec{r}_{\ell}}
D_{\ell}\vert 1\rangle_{\vec{k}}
\right\rbrace ,
\end{equation}
\begin{equation}\label{Fe_CE12}
\frac{1}{\sqrt{N}}\sum_{\ell}e^{i \vec{k}_{0}\cdot\vec{r}_{\ell}}\vert g\rangle\vert e_{\ell}\rangle\partial_{t} D_{\ell}
=
-i\left\lbrace 
\frac{1}{\sqrt{N}}\sum_{\ell}e^{i \vec{k}_{0}\cdot\vec{r}_{\ell}}\vert g\rangle\vert e_{\ell}\rangle \left[ \omega + \Delta(\vec{r}_{\ell},t) \right] D_{\ell}
+\sum_{j,\vec{u}} g_{\vec{u}}e^{i\vec{u}\cdot \vec{r}_{j}}
\vert g\rangle\vert e_{j}\rangle B_{\vec{u}}
\right\rbrace.
\end{equation}

Since $_{\vec{m}}\langle 1\vert 1\rangle_{\vec{n}}=\delta_{\vec{m}\vec{n}}$ and $\langle e_{\alpha}\vert e_{\beta}\rangle=\delta_{\alpha\beta}$, substitute $B_{\vec{u}}=b_{\vec{u}}e^{-i \nu_{\vec{u}}t}$ and $D_{\ell}=\eta e^{-i \left[ \omega t+\int^{t}_{0}\Delta(\vec{r}_{\ell},\tau)d\tau\right] }$ into Eq.~(\ref{Fe_Ck12}) and Eq.~(\ref{Fe_CE12}).
The total state vector becomes
\begin{equation}\label{Fe_totalstateb}
\vert\psi\rangle=\vert G\rangle\otimes\sum_{\vec{u}}b_{\vec{u}}(t)e^{-i \nu_{\vec{u}}t}\vert 1\rangle_{\vec{u}}+\frac{\eta(t)}{\sqrt{N}}\sum_{\ell} e^{-i \omega t }e^{i \left[\vec{k}_{0} \cdot\vec{r}_{\ell}-\int^{t}_{0}\Delta(\vec{r}_{\ell},\tau)d\tau\right]}\vert g\rangle\vert e_{\ell}\rangle\otimes\vert 0\rangle,
\end{equation}
With the assumptions of $r_{\ell}\ll R$ and constant $\varLambda$, Eq.~(\ref{Fe_Ck12},\ref{Fe_CE12},\ref{Fe_totalstateb}) become
\begin{equation}\label{Fe_totalstatec}
\vert\psi\rangle=\vert G\rangle\otimes\sum_{\vec{u}}b_{\vec{u}}(t)e^{-i \nu_{\vec{u}}t}\vert 1\rangle_{\vec{u}}+\frac{\eta(t)}{\sqrt{N}}\sum_{\ell} e^{-i (\omega+\alpha) t }e^{i \left[\vec{k}_{0}-Qt\widehat{e}_{\perp}-\int^{t}_{0}\beta(\tau)d\tau \widehat{e}_{\parallel}\right]\cdot\vec{r}_{\ell}}\vert g\rangle\vert e_{\ell}\rangle\otimes\vert 0\rangle,
\end{equation}
Let $\vec{k}_{\Delta}(t)=Qt\widehat{e}_{\perp}+\int^{t}_{0}\beta(\tau)d\tau \widehat{e}_{\parallel}$, and then one could get
\begin{eqnarray}\label{Fe_Ck22}
\nonumber
\partial_{t}b_{\vec{u}}
&=&
-\frac{i}{\sqrt{N}}g_{\vec{u}}e^{i\left( \nu_{\vec{u}}-\omega-\alpha \right) t}\eta\sum_{\ell}
e^{i(\vec{k}_{0}-\vec{k}_{\Delta}-\vec{u})\cdot \vec{r}_{\ell}}\nonumber\\
&=&
-i\frac{\sqrt{N}}{V}\left( 2\pi\right)^{3}g_{\vec{u}}e^{i\left( \nu_{\vec{u}}-\omega-\alpha \right) t}
\delta^{3}\left[ \vec{k}_{0}-\vec{k}_{\Delta}(t)-\vec{u}\right] \eta, 
\end{eqnarray}
\begin{equation}\label{Fe_CE22}
\partial_{t}\eta=-i\sqrt{N}
\sum_{\vec{u}}
g_{\vec{u}}e^{-i\left[\vec{k}_{0}-\vec{k}_{\Delta}(t)-\vec{u}\right]\cdot \vec{r}_{\ell}}e^{-i\left( \nu_{\vec{u}}-\omega-\alpha\right) t} b_{\vec{u}}.
\end{equation}
In Eq.~(\ref{Fe_Ck22}), the atomic summation (for large number density as in a solid-state sample)
\begin{equation}\label{Fe_fe22}
\sum_{\ell}e^{i\left( \vec{k}_{S}-\vec{u}\right) \cdot \vec{r}_{\ell}}=\frac{N}{V}\left( 2\pi\right)^{3}\delta^{3}\left( \vec{k}_{S}-\vec{u}\right)
\end{equation}
is used, where $V$ is the volume of the sample. Also, from Eq.~(\ref{Fe_Ck22}), we obtain
\begin{equation}\label{Fe_Ck32}
b_{\vec{u}}(t)=
-i\frac{\sqrt{N}}{V}\left( 2\pi\right)^{3}g_{\vec{u}}
\int_{0}^{t}
e^{i\left( \nu_{\vec{u}}-\omega-\alpha \right) T}
\delta^{3}\left[ \vec{k}_{0}-\vec{k}_{\Delta}(T)-\vec{u}\right] \eta(T) dT.
\end{equation}
On substituting Eq.~(\ref{Fe_Ck32}) into Eq.~(\ref{Fe_CE22}) and replacing the summation over $\vec{u}$ by an integral 
\begin{equation}
\sum_{\vec{u}}\rightarrow 
2\frac{v}{\left( 2\pi\right)^{3}}\int_{0}^{\infty} u_{r} du_{r}\int_{0}^{2\pi}d\Theta
\int_{-\infty}^{\infty}du_{z},
\end{equation} 
where $v$ is the photon volume, we obtain
\begin{eqnarray}\label{Fe_CE32}
\partial_{t}\eta &=& -\frac{\left( 2\pi\right)^{3} N}{V}
\sum_{\vec{u}}
g^{2}_{\vec{u}}e^{-i\left[\vec{k}_{0}-\vec{k}_{\Delta}(t)-\vec{u}\right]\cdot \vec{r}_{\ell}}e^{-i\left( \nu_{\vec{u}}-\omega-\alpha\right) t} \nonumber\\
&\times & 
\int_{0}^{t}
e^{i\left( \nu_{\vec{u}}-\omega-\alpha \right) T}
\delta^{3}\left[ \vec{k}_{0}-\vec{k}_{\Delta}(t)-\vec{u}\right] \eta(T) dT \nonumber\\
&=&
-\frac{2 v N}{V}
g^{2}_{\vec{k}_{S}}e^{-i\left( \nu_{\vec{u}}-\omega-\alpha\right) t} 
\int_{0}^{t}
e^{i\left( \nu_{\vec{k}_{S}}-\omega-\alpha \right) T} \eta(T) dT.
\end{eqnarray}
Here
\begin{eqnarray}
\vec{k}_{S}(t)&=& \vec{k}_{0}-\vec{k}_{\Delta}(t) \nonumber\\
              &=& \left[ k_{0}-P(t)\right] \widehat{e}_{\parallel}-t Q \widehat{e}_{\perp}
\end{eqnarray}
is the time dependent direction of scattered single photons, and $P(t)=\int^{t}_{0}\beta(\tau)d\tau$. 
Furthermore, the time dependent deflection angle $\phi(t)$ is
\begin{equation}
\phi(t)=\tan^{-1}\left( \frac{t Q }{k_{0}-P(t)}\right). 
\end{equation}
For $k_{0}\gg Qt$, one get the following formula for a small deflection angle
\begin{equation}\label{sda}
\phi(t)\approx\tan^{-1}\left( \frac{Q }{k_{0}}t\right). 
\end{equation}

\section{Energy momentum conservation}\label{semc} 
Without the longitudinal action of $P(t)$, the detuning $c|\vec{k}_{S}(t)|-\omega=c\sqrt{k_{0}^{2}+Q^{2}t^{2}}-\omega$ will exceed the finite linewidth of each considered transition.
Therefore the coherent remission may stop at some instant in time.
To make coherent remission always happen during the coherence time of some considered system, one could apply a time dependent longitudinally gradient perturbation, e.g., Stark or Zeeman fields, along the incident $\vec{k_{0}}$-direction. Under the energy momentum conservation $\nu_{\vec{k}_{S}}=\nu_{\vec{k}_{0}}$, i.e.,
\begin{equation}
\sqrt{\left[ k_{0}-P(t)\right]^{2}+Q^{2}t^{2}}=k_{0},
\end{equation}
one could find the required $P(t)$
\begin{equation}
P(t)=k_{0}-\sqrt{k_{0}^{2}-Q^{2}t^{2}},
\end{equation}
and the $\beta(t)=\partial_{t}P(t)$ is
\begin{equation}
\beta(t)=\frac{Q^{2}t}{\sqrt{k_{0}^{2}-Q^{2}t^{2}}}.
\end{equation}
Also, the deflection angle becomes
\begin{equation}
\phi(t)=\tan^{-1}\left( \frac{t Q }{\sqrt{k_{0}^{2}-Q^{2}t^{2}}}\right). 
\end{equation}
Furthermore,
\begin{equation}\label{smallbeta}
\beta(t)\simeq\frac{Q^{2}t}{k_{0}}.
\end{equation}
for small angle deflection, i.e., $k_0\gg Qt$.

In what follows, we present an example of using a rotating $^{229}$Th:CaF$_{2}$ target.
On substituting $\varLambda=2\pi\times 70$ kHz, $R=5$ mm \cite{roehlsberger2004} and $c=3\times 10^{8}$ m/s, $\alpha$ and $Q$ become 
\begin{eqnarray}
\alpha &\approx & -\frac{\omega}{2}\left( \frac{\Omega R }{c}\right)^{2}=-2.69\times 10^{-11}\omega, \\
Q &=& -5.37\times 10^{-11}\omega.
\end{eqnarray}
Considering $^{229}$Th:CaF$_{2}$ with $\omega=1.1564\times 10^{16}$ Hz and coherence time $\tau_{coh}= 1$ ms, the transverse wave number $Q\times \tau_{coh}=6214 m^{-1}$ and the deflection angle is $\tan^{-1}\left( \frac{Q\tau_{coh}}{k_{0}}\right)=3.2\times 10^{-3}$ rad. However, $\nu_{\vec{u}}-\omega=c\lvert Q\tau_{coh} \hat{e}_{r}+k_{0} \hat{e}_{z}\rvert-\omega=1.5\times 10^{8}$ Hz, which is larger than the  $^{229}$Th isomeric transition linewidth $\Gamma$ of 0.1 kHz in a CaF$_{2}$ crystal, and the deflection angle of $3.2\times 10^{-3}$ rad would not be observed.
To make the bending superradiance appear, we apply an additional magnetic gradient field $-A t z_{\ell}$ along the $\vec{k_{0}}$-direction such that an extra quantum phase evolution $\int^{t}_{0}\beta(\tau)d\tau=\frac{\mu_{n}A t^{2}}{2\hbar}$, caused by Zeeman shift, will decrease wave number in the longitudinal direction:
\begin{equation}
k_{z}=k_{0}-\frac{\mu_{n}A t^{2}}{2\hbar}.
\end{equation}
Here $\frac{\mu_{n}}{\hbar}=2.538\times 10^{7}$ Hz/T is the magnetic dipole moment of $^{229}$Th. Considering Eq.~(\ref{smallbeta})
\begin{equation}\label{energycondition} 
P(t)\simeq\frac{Q^{2}t^{2}}{2 k_{0}}=\frac{\mu_{n}A t^{2}}{2\hbar}, 
\end{equation}
we get the following formula of $A$:
\begin{equation}
A=\frac{\hbar Q^{2}}{k_{0} \mu_{n}}\sim 2.52\times 10^{-7}  \ (\bf{\frac{T}{s\cdot m}}).
\end{equation}

\section{Photon counts on the position sensitive detector} 
\subsection{For a narrow band excitation} 
To get the behavior of photon count on the position sensitive detector, one has to solve Eq.~(\ref{Fe_CE32}), Let
\begin{eqnarray}
\chi(t) &=&
\int_{0}^{t}
e^{i\left( \nu_{\vec{k}_{S}}-\omega-\alpha \right) \tau} \eta(\tau) d\tau, \label{xi}\\
\partial_{t}\chi(t) &=&
e^{i\left( \nu_{\vec{k}_{S}}-\omega-\alpha \right) t} \eta(t), \label{dxi}\\
\partial^{2}_{t}\chi(t) &=&
i\left( \nu_{\vec{k}_{S}}-\omega-\alpha +t\partial_{t}\nu_{\vec{k}_{S}}\right)e^{i\left( \nu_{\vec{k}_{S}}-\omega-\alpha \right) t} \eta(t) \nonumber
+e^{i\left( \nu_{\vec{k}_{S}}-\omega-\alpha \right) t} \partial_{t}\eta(t)\\
&=&
i\left( \nu_{\vec{k}_{S}}-\omega-\alpha +t\partial_{t}\nu_{\vec{k}_{S}}\right)\partial_{t}\chi(t)
+e^{i\left( \nu_{\vec{k}_{S}}-\omega-\alpha \right) t} \partial_{t}\eta(t), \label{ddxi}
\end{eqnarray}
and substitute Eq.~(\ref{Fe_CE32}) and Eq.~(\ref{dxi}) into Eq.~(\ref{ddxi}), we get
\begin{equation}
\partial^{2}_{t}\chi(t)-i\left( \nu_{\vec{k}_{S}}-\omega-\alpha +t\partial_{t}\nu_{\vec{k}_{S}}\right)\partial_{t}\chi(t)+\frac{2 v N}{V}
g^{2}_{\vec{k}_{S}}\chi(t)=0.
\end{equation}
By making $v=V$ since a single photon is absorbed by the whole ensemble and using 
\begin{equation}
\Omega^{2}=2 g_{\vec{k}_{S}}^{2}=\frac{\nu_{\vec{k}_{S}}}{\hbar\epsilon_{0}V}\varrho_{eg}^{2},
\end{equation}
and the energy momentum conservation
\begin{equation}\label{emc}
\nu_{\vec{k}_{S}}=\nu_{\vec{k}_{0}}=\omega,
\end{equation}
which results in $\partial_{t}\nu_{\vec{k}_{S}}=0$. The equation of motion of $\chi$ then reads
\begin{equation}
\partial^{2}_{t}\chi(t)+i \alpha\partial_{t}\chi(t)
+N\Omega^2\chi(t)=0,
\end{equation}
and initial conditions according to Eq.~(\ref{ic}, \ref{xi}, \ref{dxi})
\begin{eqnarray}
\chi(0)=0,\\
\partial_{t}\chi(t)\rvert_{t=0}=1.
\end{eqnarray}
We get 
\begin{eqnarray}
\chi(t)&=&\frac{2e^{-i\frac{\alpha}{2} t}}{\sqrt{\alpha^{2}+4N\Omega^{2}}}\sin\left( \frac{\sqrt{\alpha^{2}+4N\Omega^{2}}}{2}t\right),\\
\eta(t)&=&\frac{-i\alpha e^{i\frac{\alpha}{2} t}}{\sqrt{\alpha^{2}+4N\Omega^{2}}}\sin\left( \frac{\sqrt{\alpha^{2}+4N\Omega^{2}}}{2}t\right)
+
e^{i\frac{\alpha}{2} t}\cos\left( \frac{\sqrt{\alpha^{2}+4N\Omega^{2}}}{2}t\right), \\
b_{\vec{u}}(t)&=&
-i\frac{\sqrt{N}}{V}\left( 2\pi\right)^{3}g_{\vec{u}}
\int_{0}^{t}
e^{-i\alpha T}
\delta^{3}\left[ \vec{k}_{S}(T)-\vec{u}\right] \eta(T) dT. \label{be}
\end{eqnarray}
Here $\delta^{3}\left[ \vec{k}_{S}(T)-\vec{u}\right]$ does not make any time constrain on $\eta(T)$, but on $\vec{u}$. Therefore Eq.~(\ref{be}) becomes
\begin{eqnarray}
b_{\vec{k_{S}}(t)}(t)&=&
-i\frac{\sqrt{N}}{V}\left( 2\pi\right)^{3}g_{\vec{k}_{S}}
\int_{0}^{t}
e^{-i\alpha T}\eta(T) dT. \nonumber\\
&=&-\frac{2i}{\sqrt{\alpha^{2}+4N\Omega^{2}}}e^{-i\frac{\alpha}{2} t}\sin\left( \frac{\sqrt{\alpha^{2}+4N\Omega^{2}}}{2}t\right)
\end{eqnarray}
Furthermore, for $\alpha\ll 2\sqrt{N}\Omega$ we get
\begin{equation}\label{eqosci1}
\eta(t)=e^{i\frac{\alpha}{2} t}\cos\left( \sqrt{N}\Omega t\right),
\end{equation}
\begin{equation}\label{eqosci2}
b_{\vec{k}_{S}}(t)=-\frac{i}{\sqrt{N}\Omega}e^{-i\frac{\alpha}{2} t}\sin\left( \sqrt{N}\Omega t\right), 
\end{equation}
where $\vec{k}_{S}(t)=\left[ k_{0}-P(t)\right] \widehat{e}_{\parallel}-t Q \widehat{e}_{\perp}$. Count numbers behavior is
\begin{eqnarray}
r(t) &=& R+\frac{h}{2}+D\tan\left( \phi_{c}(t)\right), \\
\Theta(t) &=& \varLambda t,\\
I(t) &\propto& \sin^2\left( \sqrt{N}\Omega t\right),
\end{eqnarray}
i.e., at each $(r,\Theta)$ on the detector the count number $I(t)$ is proportional to $\sin^2\left( \sqrt{N}\Omega t\right)$. Here $R$, $h$, $D$ and $\varLambda$ are the rotor radius, target height, distance between a target and  a detector and rotating angular frequency of a rotor.

However, if the ensemble volume $V$ is finite such that the atomic summation becomes \cite{svidzinsky2008}
\begin{equation}\label{as2}
\sum_{\ell}e^{i\left( \vec{k}_{S}-\vec{k}\right) \cdot \vec{r}_{\ell}}=\frac{\sin\left(|\vec{k}_{S}-\vec{k}|\sqrt[3]{V} \right)}{|\vec{k}_{S}-\vec{k}|^3}-\frac{\sqrt[3]{V}\cos\left(|\vec{k}_{S}-\vec{k}|\sqrt[3]{V} \right)}{|\vec{k}_{S}-\vec{k}|^2},
\end{equation}
$I(t)$ turns into \cite{svidzinsky2008}

\begin{equation}
I(t) \propto \left\{ 
\begin{array}{ccc}
  \sin^2\left( \sqrt{N}\Omega t\right)exp\left( -\frac{3c}{4\sqrt[3]{V}}t\right) & {\bf for} & \frac{\sqrt[3]{V}}{c}\gg\frac{1}{\sqrt{N}\Omega} \\
  exp\left( -\frac{3N\Omega^{2}\sqrt[3]{V}}{4c}t\right) & {\bf for} & \frac{\sqrt[3]{V}}{c}\ll\frac{1}{\sqrt{N}\Omega} 
\end{array} 
\right.
\end{equation}
Also, according to Eq.~(\ref{as2}) there is a finite divergence angle $\Delta\phi$ of a remitted photon 
\begin{equation}\label{eq58}
\Delta\phi=\frac{\pi}{k_{0}\sqrt[3]{V}}.
\end{equation}
Therefore together with the relativistic photon deflection, there will be a diffraction drag caused by the finite size effect on the detector. However, $\Delta\phi$ would be negligible in hard x-ray region whose wavelength is shorter than $10^{-10}$ m and much smaller than any size of an used sample.

\subsection{For a broadband excitation}
In above calculation, we demonstrate so-called cooperative spontaneous emission \cite{scully2006} following the decay of the state $\vert E \rangle$ excited by a single color photon of $\vert 1\rangle_{\vec{k_0}}$.
Once the bandwidth of the incident photon is much larger than the linewidth of the interacting transition (this is the typical case for nuclear coherent scattering of x-ray \cite{roehlsberger2004}), the dispersion of the pulse propagation must be considered. To deal with such case  for the perturbation region,  in the field of nuclear coherent scattering \cite{roehlsberger2004} a semiclassical approach, namely Optical-Bloch equation \cite{scully1997}, is typically used
\begin{equation}\label{TM_twolevelp23}
\partial_{t}\rho_{21} =  -\left(\frac{\Gamma}{2}+i\Delta \right)\rho_{21}
+\frac{i}{2}\Omega_{p}.
\end{equation}
Here $\rho_{21}=C_{1}C_{1}^{\ast}$
for a state vector of a two-level particle $\vert\psi\rangle=C_{1}(t)\vert 1\rangle+C_{2}(t)\vert 2\rangle$.
By making $\partial_{t}\rho_{21} =0$, one could get the steady solution of Eq.~(\ref{TM_twolevelp23})
\begin{equation}
\rho_{21} = \frac{2\Delta\Omega_{p}}{\Gamma^2+4\Delta^2}+i \frac{2\Gamma\Omega_{p}}{\Gamma^2+4\Delta^2}.
\end{equation}
The field susceptibility is \cite{Zhu2013}
\begin{equation}
 \chi_{p} = \frac{\wp\vert\vec{d}\vert^{2}}{\hbar\epsilon_{0}\Omega_{p}}\rho_{21},
\end{equation}
where $\wp$, $\vec{d}$ and $\epsilon_{0}$ are particle density, transition dipole moment and  vacuum permittivity, respectively.
By substituting the so called on-resonance cross section
\begin{equation}
\sigma=\frac{4\pi\vert\vec{d}\vert^{2}}{\hbar\epsilon_{0}\lambda\Gamma},
\end{equation}
into $\chi_{p}$, we get
\begin{eqnarray}
\chi_{p} &=& \frac{\wp\sigma\Gamma}{2 k_{0}\Omega_{p}}\rho_{21}, \\
         &=& \frac{\wp\sigma\Gamma}{k_{0}} \left(\frac{\Delta}{\Gamma^2+4\Delta^2}+\frac{i\Gamma}{\Gamma^2+4\Delta^2}\right),
\end{eqnarray}
and the index of refraction $n_p=\sqrt{1+\chi_{p}}=n_p'+in_p''$.

\subsubsection{Deflection angle}
To calculate the deflection angle, we invoke the well-known geometrical optics differential equation in vector form \cite{Zhu2013,Dressel2009}
\begin{equation}\label{gode}
\frac{d}{ds}\left[n_p'(r,z)\frac{d\vec{\Upsilon}}{ds} \right]=\bigtriangledown n_p', 
\end{equation}
where $ds=\sqrt{dr^2+dz^2}$ and $\vec{\Upsilon}=(r,z)$. The explicit form of Eq.~(\ref{gode}) is
\begin{eqnarray}
\frac{d}{ds}\left[n_p'(r,z)\frac{dr}{ds} \right]&=&\frac{\partial n_p'}{\partial r}\hat{e}_{\perp}, \label{longi}\\
\frac{d}{ds}\left[n_p'(r,z)\frac{dz}{ds} \right]&=&\frac{\partial n_p'}{\partial z}\hat{e}_{\parallel}, 
\end{eqnarray}
where
\begin{eqnarray}
n_p'(r,z) &\approx& 1+\frac{\wp\sigma\Gamma}{2 k_{0}} \left(\frac{\Delta}{\Gamma^2+4\Delta^2}\right),\\
\frac{\partial n_p'}{\partial r} &=& \frac{\wp\sigma\Gamma}{2 k_{0}}\frac{Q\left(\Gamma^2-4\Delta^2\right)}{\left(\Gamma^2+4\Delta^2\right)^2}\approx\frac{\wp\sigma Q}{k_{0}\Gamma}.
\end{eqnarray}
Also, for a small deflection $ds\approx dz$ \cite{Zhu2013}, and so from Eq.~(\ref{longi}) one could obtain
\begin{eqnarray}
 n_p'(r,z)\frac{dr}{dz} &=& \frac{\wp\sigma Q}{k_{0}\Gamma}\int_{0}^{L}dz\\
                              &=& \frac{Q}{k_{0}\Gamma}\wp\sigma L.\label{da1}
\end{eqnarray}
If $ n_p'\approx 1$, Eq.~(\ref{da1}) becomes \cite{Dressel2009,Zhu2013}
\begin{equation}
\tan\phi = \frac{dr}{dz} = \wp\sigma L\frac{Q}{k_{0}\Gamma},
\end{equation}
where the maximum deflection angle is 
\begin{equation}\label{da2}
\phi = \tan^{-1}\left(\wp\sigma L\frac{Q}{k_{0}\Gamma}\right).
\end{equation}
This confirms Eq.~(\ref{sda}) when an optical depth $\wp\sigma L=1$ and $t=1/\Gamma$, i.e., the maximum deflection angle within the coherence time of the interacting transition in an optical thin medium. On the other hand, Eq.~(\ref{da2}) shows that an optical thick medium will enhance the coherent deflection mechanism \cite{Dressel2009,Zhu2013}.

\subsubsection{Count behavior on a position-sensitive detector}
To calculate the photon count behavior of nuclear coherent scattering, one can solve Eq.~(\ref{TM_twolevelp23}) together with the wave equation \cite{scully1997, Crisp1970, Shvydko1999}
\begin{equation}
\frac{1}{c}\partial_{t}\Omega_p+\partial_{s}\Omega_p=i\eta\rho_{21}\label{TM_maxwell},
\end{equation}
where $\partial_{s}\approx \partial_{z}$ for a small deflection \cite{Zhu2013}, and $\eta=\frac{\Gamma\wp\sigma}{2}$.
Moreover, since the pulse duration of the incident x-ray is much shorter than the lifetime of the nuclear excited state, the following boundary and initial conditions of $\Omega_p(t,z)$ can be used \cite{Crisp1970, Shvydko1999}
\begin{eqnarray}
\Omega_p(t,0)  &=&  \delta(t), \\
\Omega_p(0,z)  &=&  0.
\end{eqnarray}
The solution of $\Omega_p(t,z)$ for Eq.~(\ref{TM_twolevelp23},\ref{TM_maxwell}) is \cite{Crisp1970, Shvydko1999}
\begin{equation}\label{bessel}
\Omega_{p}(t,z) =
\delta(t)
- \eta s(t,z)\frac{J_{1}\left[  2\sqrt{ \frac{\eta s(t,z) }{2} t}\,\right]  }{4\sqrt{\frac{\eta s(t,z)}{2}t }}
e^{-\frac{\Gamma}{2}t   },
\end{equation}
where $\phi(t,z)=\tan^{-1}\left(\wp\sigma z\frac{Q}{k_{0}}t\right)$ and $J_{1}$ is the Bessel function of first king of first order. 
Furthermore, $s(t,z)$ is the length of the deflected x-ray trajectory at longitudinal position $z$ in a crystal and reads
\begin{eqnarray}
s(t,z) &=& \int_0^z\sec\left[\phi(t,z)\right]dz', \nonumber\\
       &=& \int_0^z\sqrt{1+\left(\wp\sigma z'\frac{Q}{k_{0}}t\right)^2} dz', \nonumber\\
       &=& \frac{z}{2}\sqrt{1+\left(\wp\sigma z\frac{Q}{k_{0}}t\right)^2}+\frac{k_0}{2\wp\sigma Q t}\sinh^{-1}\left(\wp\sigma z\frac{Q}{k_{0}}t\right).
\end{eqnarray}

The photon count count registered by a detector is $I(t)\propto\vert\Omega_{p}(t,L)\vert^2$ at a deflection angle $\phi(t,L)$.

\section{Minimum length scale for gravitational deflections}
For finding out the minimum length scale for gravitational deflection of superradiance, one could invoke Eq.~(\ref{eq58}) and $\phi_g(\tau_{coh})>\Delta\phi$, i.e., the maximum deflection angle is larger than the divergence in deflection measurements. By constraining $\Delta\phi=\phi_g(\tau_{coh})/10$ and using Eq.~(\ref{eq58}), one could get
\begin{equation}
\sqrt[3]{V}=\frac{10\pi}{k_{0}\phi_g(\tau_{coh})}.
\end{equation}
Considering $^{109}$Ag with $\phi_g(\tau_{coh})=1.1\times 10^{-4}$ degree and $k_0=4.46\times 10^{11}$ m$^{-1}$, the minimum length scale $\sqrt[3]{V}=3.7\times 10^{-5}$ m.

\section{Comparison of light deflection with radial motion of a particle in curved space-time of general relativity}

\noindent The light deflection angle of free propagating particles with the speed of light can be computed by making use of the solution
\begin{equation}
 u = \frac{\sin \varphi}{b} + \frac{3 G M_E}{2 c^2 b^2} \left( 1 + \frac{1}{3} \cos 2 \varphi \right) \label{eq:u}
\end{equation}
of the inverse radius $u=1/r$ \cite{hobson2006}. In this solution, the coordinates are implicitly choosen, such that the particle comes from $\varphi \approx 0$ and goes to $\varphi \approx \pi$. If the second term in \eqref{eq:u} of the gravitational perturbation was zero, the particle would come exactly from $\varphi = 0$ go exactly to $\varphi = \pi$, but a non-vanishing gravitational potential yields a slight deviation, from which the gravitational deflection is computed (see reference \cite{hobson2006}). However, we are interested in the radial acceleration of the particle, when it is passing by the gravitational center and is nearest to it. In this case, the propagation direction of the particle is perpendicular to the radius vector, which happens at the point, at which $\varphi=\pi/2$. The radial acceleration is given by the doubled time-derivative of the radius $r$ and can be related to the solution of $u$ by
\begin{subequations}
\begin{align}
 \dot r &= \frac{\partial}{\partial t} r = \frac{\partial}{\partial t} \frac{1}{u} = - \frac{1}{u^2} \dot u \,,\\
 \ddot r &= \frac{\partial^2}{\partial t^2} r = \frac{\partial}{\partial t} \left( - \frac{1}{u^2} \dot u \right) = \frac{2}{u^3} {\dot u}^2 - \frac{1}{u^2} \ddot u\,.
\end{align}
\end{subequations}
The first and second time-derivative of $u$  evaluate as
\begin{subequations}
\begin{align}
 \dot u &= \frac{\partial}{\partial t} u = \frac{1}{b} \dot \varphi \cos \varphi - \frac{G M_E}{c^2 b^2} \dot \varphi \sin 2 \varphi \,,\\
 \ddot u &= \frac{\partial}{\partial t} \dot u = \frac{1}{b} \left( - {\dot \varphi}^2 \sin \varphi + \ddot \varphi \cos \varphi \right) - \frac{G M_E}{c^2 b^2} \left( 2 {\dot \varphi}^2 \cos 2 \varphi + \ddot \varphi \sin 2 \varphi \right)\,.
\end{align}
\end{subequations}
$\dot u$ is zero at the angle $\varphi = \pi/2$ and therewith $\dot r$ is zero too. For an evaluation of $\ddot u$, the first and second time-derivatives of $\varphi$ have to be computed. The angular velocity $\dot \varphi$ of the particle is implied by the line element \eqref{eq:lineelement} and with $\dot r = 0$ and the property that light is a null geodesic ($dt^{\prime 2}=0$) one obtains the condition
\begin{equation}
 0 = \left(1 - \frac{r_s}{r}\right) c^2 dt^2 - r^2 d\varphi^2\,.
\end{equation}
Dividing by $dt$ yields
\begin{equation}
 \dot \varphi =  \frac{c}{r} \sqrt{1 - \frac{r_s}{r}}\,.
\end{equation}
The angular acceleration
\begin{equation}
 \ddot \varphi = - \frac{c}{r^2} \sqrt{1 - \frac{r_s}{r}} \, \dot r + \frac{c}{r} \left( \frac{1}{2} \sqrt{1 - \frac{r_s}{r}}^{-1} \right) \frac{r_s}{r^2} \dot r
\end{equation}
vanishes at $\varphi = \pi/2$, because of $\dot r = 0$. In summary we have
\begin{subequations}
\begin{align}
 \varphi &= \frac{\pi}{2} \\
 \left. \dot \varphi \right|_{\varphi=\frac{\pi}{2}} &= \frac{c}{r} \sqrt{1 - \frac{r_s}{r}} \label{eq:dot_phi}\\
 \left. \ddot \varphi \right|_{\varphi=\frac{\pi}{2}} &= 0
\end{align}
\end{subequations}
Inserting this into $u$ and it's derivatives yields
\begin{subequations}
\begin{align}
 \left. u \right|_{\varphi=\frac{\pi}{2}} &= \frac{1}{b} + \frac{G M_E}{c^2 b^2} \,,\\
 \left. \dot u \right|_{\varphi=\frac{\pi}{2}} &= 0 \,,\\
 \left. \ddot u \right|_{\varphi=\frac{\pi}{2}} &= - \frac{{\dot \varphi}^2}{b} + 2 \frac{G M_E}{c^2 b^2} {\dot \varphi}^2\,.
\end{align}
\end{subequations}
Thus the radius and it's derivatives turn into
\begin{subequations}
\begin{align}
 \left. r \right|_{\varphi=\frac{\pi}{2}} &= \left( \frac{1}{b} + \frac{G M_E}{c^2 b^2} \right) \,,\\
 \left. \dot r \right|_{\varphi=\frac{\pi}{2}} &= 0 \,,\\
 \left. \ddot r \right|_{\varphi=\frac{\pi}{2}} &= - r^2 \left ( - \frac{{\dot \varphi}^2}{b} + 2 \frac{G M_E}{c^2 b^2} {\dot \varphi}^2 \right)\,. \label{eq:ddot_r}
\end{align}
\end{subequations}
With equation \eqref{eq:ddot_r} we have derived an expression for the radial acceleration at $\varphi = \pi/2$ without applying any approximations. We now introduce approximations, to compare our result with Newtonian physics and with our light deflection equation (4). Since the influence by gravity is assumed to be a small perturbation, $b$ turns into $1/u$ (see Eq. (10.10) of \cite{hobson2006}), and since $u$ is the inverse radius, we have
\begin{equation}
 \frac{1}{u} \approx b \approx r_E
\end{equation}
at earth surface. Furthermore, at earth the radius, the fraction $r_s/r_E$ is $1.3\times10^{-9}$ and with the approximation
\begin{equation}
 \left( 1 - \frac{r_s}{r} \right) \approx 1\,,\label{eq:small_rs_approximation}
\end{equation}
the angular velocity $\dot \varphi$ of equation \eqref{eq:dot_phi} turns into
\begin{equation}
 \dot \varphi \approx \frac{c}{r_E}\,.
\end{equation}
Finally, the radial acceleration \eqref{eq:ddot_r} simplifies to
\begin{equation}
 \ddot r \approx \frac{c^2}{r_E} - 2 \frac{G M_E}{r_E^2} \,. \label{eq:GR-acceleration}
\end{equation}

\noindent The first term $c^2/r_E$ is a relict which originates from the usage of spherical coordinates. If one considers the four vector parameterization of a particle with speed of light $(ct,-c t,r_E,0)$ in flat, cartesian space-time, with corresponding spherical coordinates
\begin{subequations}
\begin{align}
 r &= \sqrt{r_E^2 + c^2 t^2} \\
 \theta &= \frac{\pi}{2} \\
 \varphi + \frac{\pi}{2} &= \tan^{-1}\left(\frac{ct}{r_E}\right)\,.
\end{align}
\end{subequations}
The doubled time-derivative of the radius at time $t=0$ (ie. at $\varphi=\pi/2$)
\begin{equation}
 \left.\frac{\partial^2 r}{\partial t^2} \right|_{t=0} = \left. \frac{c^2 t^2}{\sqrt{r_E^2 + c^2 t^2}^2} \right|_{t=0} = \frac{c^2}{r_E}
\end{equation}
yields exactly the first term of Eq. \eqref{eq:GR-acceleration}, emphasizing that it is only an artefact of straight lines in spherical coordinates.

\noindent Eq. \eqref{eq:GR-acceleration} demonstates, that light is deflected twice as much as one would expect it from a classical massive particle in Newtonian theory. This discrepancy in light deflection is known as the first experimentally verified prediction of Einstein's theory of general relativity \cite{hobson2006}. According to this radial acceleration a freely propagating photon acquires the velocity change
\begin{equation}
 2 \frac{G M_E}{r_E^2} \Delta t \label{eq:velocity_change}
\end{equation}
towards the earth radial direction after some infinitesimal time $\Delta t$. We compare this with the gravitational deflection formula (4) in the main text. The inverse tangent can be approximated linearly $\tan^{-1} x \approx x$ for small $x$ and together with the approximation \eqref{eq:small_rs_approximation} the expression reads
\begin{equation}
 \phi_g(\Delta t) \approx \frac{G M_E}{c r_E^2} \Delta t\,.
\end{equation}
If the four vector $(c,-c,0,0)^T$ of light propagating in $x$-direction is rotated by this infinitesimal deflection angle $\phi_g(\Delta t)$ in $r$-direction, the factor $c$ in the denominator lifts away and we obtain a velocity change, which is by a factor of 2 smaller as in equation \eqref{eq:velocity_change}.

Therefore, we conclude that the superradiant light, which is stored as coherent excitation in the crystal does not acquire the same acceleration as freely propagating light in a gravitational field. Instead, we conclude that the superradiant light is deflected like a classical massive particle in Newtons theory.

\section{Comparison of special-relativistic superradiant deflection with nuclear lighthouse effect}
The essential difference between our rotating setup and the nuclear lighthouse effect is illustrated in Fig.~\ref{fig1}. 
The direction of the centrifugal acceleration and the $\vec{k}_{0}$ of incident x-rays are perpendicular (parallel) to each other in our rotating crystal setup (the nuclear lighthouse setup). 
Therefore, the centrifugal acceleration will not deflect photons in nuclear lighthouse effect, but the x-ray wave vector changes due to the mechanical motion of the nuclear crystal (like rotating a laser pointer along an axis perpendicular to its cavity axis). On the other hand, in our case, if one takes only the crystal's mechanical motion  into account without any influence of special relativity, the registered superradiant photons will only form a ring on a detector. The illustrated spiral pattern on a image plate is only possible when the action of special-relativistic time dilation is considered. This gives the major difference between our special-relativistic superradiant deflection and the nuclear lighthouse effect.
\begin{figure}[h]
\begin{center}
  \includegraphics[width=1\textwidth]{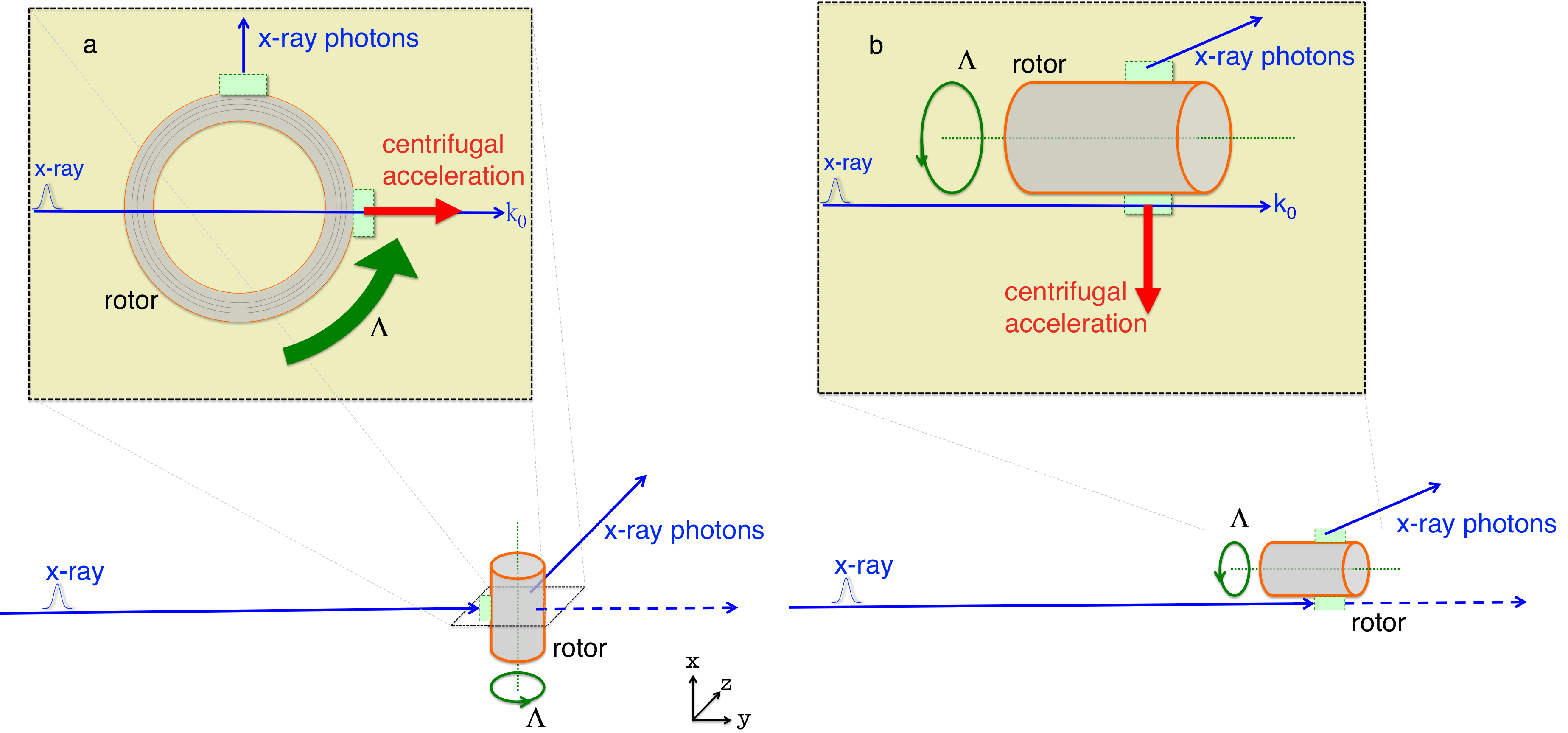}
  \caption{\label{fig1} Comparison of the superradiant deflection setup with nuclear lighthouse setup. The direction of the centrifugal acceleration is parallel (perpendicular) to the wave vector $\vec{k}_{0}$ of the incident x-rays in (\textbf{a}) the nuclear lighthouse scheme [(\textbf{b}) the special-relativistic superradiant deflection setup]. The blue thin (red thick) arrows illustrate the wave vectors of x-rays (centrifugal acceleration). The light green rectangles depict snapshots of the rotating nuclear crystal attached on the surface of the rotor.
}
\end{center}
\end{figure}

\section{Possible implementation issues}
In x-ray scattering experiments, one could use so-called Bose-Hart geometry  to achieve sub-$\mu$rad angular resolution for ultra small angle x-ray scattering (USAXS) \cite{Tamasaku2014}.
For using a rotating crystal, the wobble of a used rotor and the deformation of the crystal may affect the measurement of photon deflection. To discuss the wobble problem, we have checked the experimental data from nuclear lighthouse effect, see, e.g., Fig. 6.6 of Ref. \cite{roehlsberger2004}. The trajectory of the registered photons on the image plate shows a straight line with a resolution of at least 10$^{-4}$ rad, and so the wobble of a used rotor might be negligible comparing to some of our calculated deflection angles. As for the crystal deformation, we have also checked the experimental results of the original nuclear lighthouse paper \cite{roehlsberger2000}. A nearly perfect mapping from a time spectrum to an angular spectrum (see Fig. 4 of \cite{roehlsberger2000}) suggests that a possible deformation of crystal causes no additional deflection of x-rays. Even if one assumes that the rotation of the crystal excites different photon modes, one could still perform a nuclear x-ray scattering experiment after all phonon excitations decayed, where the lifetime of phonons is typically much shorter than 1 second..



\section{Summary}
\begin{itemize} 
  \item The time dependent deflection of superradiant emission caused by special and general relativity: this effect can be revealed by an inspection of the collective state
\begin{equation}
\vert E \rangle=\frac{1}{\sqrt{N}}\sum_{\ell}e^{i \vec{k}_{S}(t) \cdot\vec{r}_{\ell}}\vert g\rangle\vert e_{\ell}\rangle.
\end{equation}
The collective excitation is excited by one photon absorption with the wave vector $\vec{k_{0}}$, and emits a photon later in the direction of $\vec{k}_{S}(t)=\left[ k_{0}-P(t)\right] \widehat{e}_{\parallel}-t Q \widehat{e}_{\perp}$, where $Q$ is induced by an inhomogeneous speed in a rotating system or by the Earth's gravity.
  \item According to Eq.~(\ref{da2}), the relativistic deflection may be enhanced when using an optically thick crystal \cite{Duan2001}.
  \item Collective oscillation: in free space the spontaneous decay behavior of a single excited particle can be derived from the Weisskopf-Wigner theory. With the same theory a large  number of particles in the collective excitation state results in the collective oscillation described by Eq.~(\ref{eqosci1}, \ref{eqosci2}). The  oscillation can be understood as the emitted photon being re-absorbed and re-emitted in the ensemble \cite{svidzinsky2008}, i.e., multiple scattering. For a broadband excitation, the dispersion or multiple scattering \cite{Shvydko1999} results in the Bessel function Eq.~(\ref{bessel}) behavior.
  \item Here, we summarize useful formulas in Table~\ref{formulatable}.
\begin{table*}[h]
\vspace{-0.4cm}
\caption{\label{formulatable}
Summary of useful formulas.
}
\center{
\begin{tabular}{clcc}
\hline
Symbol           & Explanation          & In gravity  & In a rotating crystal        \\
\hline \\
$Q$       & $\nabla\omega'\cdot\hat{e}_{\perp}$  & $\frac{\omega r_s}{2 r_E^2 \sqrt{1 - \frac{r_s}{r_E}}}$   & $-\frac{1}{c^2}\frac{R \varLambda^2 \omega}{\sqrt{1- \frac{\varLambda^2 R^2}{c^2}}}$            \\ 
$\beta(t)$& $\nabla\omega'\cdot\hat{e}_{\parallel}$  & $\frac{c^2 r_s^2 k_0 t}{2r_E^3(r_E-r_s)\sqrt{4+\frac{c^2 t^2}{r_E (r_s-r_E)}}}$      &  $\frac{k_0 R^2 \varLambda^4 t}{(c^2- \varLambda^2 R^2)\sqrt{\frac{c^2-\varLambda^2 R^2 (1+\varLambda^2 t^2)}{c^2- \varLambda^2 R^2}}}$      \\ 
$A(t)$    & time derivative of Zeeman field gradient   & $\frac{c^2 r_s^2 \hbar k_0}{4\mu_n r_E^3(r_E-r_s)}$              & $\frac{\hbar k_{0} R^2 \varLambda^4}{\mu_{n}(c^2-\varLambda^2 R^2)}$ \\
$\phi(t)$ & deflection angle                           & $\tan^{-1}\left[ \frac{\frac{c r_s t}{r_E^2 \sqrt{1-\frac{r_s}{r_E}}}}{\sqrt{4+\frac{c^2 r_s^2 t^2}{r_E^3 (r_s-r_E)}}} \right]$&  $\tan^{-1}\left[\frac{R\varLambda^2 t}{c^2-\varLambda^2 R^2-R^2 \varLambda^4 t^2}   \right]$ \\
$\phi(t)$ & small deflection angle for $k_{0}\gg Qt$ & $\tan^{-1}\left[ \frac{G M_E t}{c r_E^2 \sqrt{1 - \frac{2 G M_E}{c^2 r_E}}} \right]$&  $\tan^{-1}\left[ \frac{R \varLambda^2 t}{\sqrt{c^{2}- \varLambda^2 R^2}}\right]$ \\
\hline
\end{tabular}
}
\end{table*}
\end{itemize}

\end{document}